\documentclass{article}
\usepackage{spconf,amsmath}
\usepackage{cite}
\usepackage{amsmath,amssymb,mathtools}
\usepackage{array}
\usepackage{stfloats}
\usepackage{algorithm,algpseudocode}
\usepackage[pdftex]{graphicx}
\usepackage{lipsum}
\usepackage{multirow}
\usepackage{textcomp}
\usepackage{url}
\usepackage{color}
\algnewcommand\INPUT{\item[\textbf{Input:}]}%
\algnewcommand\OUTPUT{\item[\textbf{Output:}]}%
\begin{document}
% Title.
% ------
\title{SINGLE IMAGE SUPER-RESOLUTION OF NOISY 3D DENTAL CT IMAGES USING TUCKER DECOMPOSITION}
%
% Single address.
% ---------------
\name{J. Hatvani$^{\dagger \star}$ \qquad A. Basarab$^{\dagger}$ \qquad J. Michetti $^{\dagger}$ \qquad M. Gy\"ongy $^{\star}$ \qquad D. Kouam\'e $^{\dagger}$ 
\thanks{This research has been partially supported by  P\'azm\'any University KAP19-20.}
}

\address{ $^{\dagger}$Universit\'e Paul Sabatier Toulouse 3 \qquad\qquad\qquad\quad $^{\star}$ P\'azm\'any P\'eter Catholic Unviersity\\
	Institut de Recherche en Informatique de Toulouse \qquad Faculty of Information Technology and Bionics   \\
	118 r. de Narbonne, F-31062 Toulouse \qquad\qquad\qquad 50/a Pr\'ater utca, H-1083 Budapest \\
%    $^{\dagger}$ Affiliation Number Two
    }
%
% For example:
% ------------
%\address{School\\
%	Department\\
%	Address}
%
% Two addresses (uncomment and modify for two-address case).
% ----------------------------------------------------------
%\twoauthors
%  {A. Author-one, B. Author-two\sthanks{Thanks to XYZ agency for funding.}}
%	{School A-B\\
%	Department A-B\\
%	Address A-B}
%  {C. Author-three, D. Author-four\sthanks{The fourth author performed the work
%	while at ...}}
%	{School C-D\\
%	Department C-D\\
%	Address C-D}
%

%\ninept
%
\maketitle
\begin{abstract}
Tensor decomposition has proven to be a strong tool in various 3D image processing tasks such as denoising and super-resolution. 
%In different  works,  the canonical polyadic decomposition (CPD) has been  investigated.
In this context, we recently proposed a canonical
polyadic decomposition (CPD) based algorithm for  single image super-resolution (SISR). The algorithm has shown to be  an order of magnitude faster than popular optimization-based techniques. In this work, we investigated the added value brought by Tucker decomposition.
While CPD allows a joint implementation of the denoising and deconvolution steps of the SISR model, with Tucker decomposition the denoising is realized first, followed by deconvolution. This way the ill-posedness of the deconvolution caused by noise is partially mitigated.
The results achieved using the two different tensor decomposition techniques were compared, and the robustness against noise was investigated. For validation, we used dental images. The superiority of the proposed method is shown in terms of peak signal-to-ratio, structural similarity index, the canal segmentation accuracy, and runtime.

\end{abstract}
\begin{keywords}
3D single image super-resolution, tensor factorization, Tucker decomposition, dental CT 
\end{keywords}
\section{Introduction}
\label{sec:intro}

Single image super-resolution (SISR) techniques aim to improve the observed image without further measurements, offering a safe and cost-effective reconstruction method. This is especially true in medical imaging, where additional radiation is to be avoided. For instance in dentistry, the position and structure of the tooth canal is determined using cone-beam computed tomography (CBCT), and is outstandingly important in case of routine root canal treatments, where the current success rate of 60-85\% \cite{ng2007outcome,eriksen2002endodontic} could be improved with SISR.

The super-resolution problem assumes an image degradation model where the high-resolution (HR) image is convolved by a blurring kernel, downsampled, and corrupted by additive noise, resulting in the low-resolution (LR) image. The problem is further complicated by its dimensionality, as the processing of 3D images is computationally more challenging. This ill-posed inverse problem was addressed in the literature by minimization techniques using different well-known regularizers \cite{toma2014total,shi2015lrtv,zhang2018limited,manjon2010non}, or deep learning techniques \cite{Cengiz17,Zhang17,hatvani2018deep}. While the first group of algorithms is computationally inefficient as a result of large matrix operations,  deep learning techniques require large training sets, which are usually difficult to obtain for medical applications.

In our previous work \cite{hatvani2018tensor} an SISR technique (TF-SISR) using canonical polyadic decomposition (CPD) was proposed, adapting the idea of \cite{Kanatsoulis18} used in multispectral-hyperspectral image fusion. 
In that work promising results were obtained for high SNR. In this paper an SISR method is proposed based on a different tensor factorization, the Tucker decomposition (TD), already used in multi- and hyper-spectral imaging \cite{prevost2019hyperspectral}. The proposed algorithm, denoted by TD-SISR, is shown to be more accurate and more robust to noise than the previous TF-SISR method. 
%The question of noise was not investigated in detail earlier. Addressing this question the Tucker decomposition (TD) was found to be a promising method, also used previously in the multispectral-hyperspectral problem \cite{prevost2019hyperspectral}. Thus, this paper shows the potentials of TD in the SISR problems, TD-SISR.

In the next section, the two tensor decomposition techniques are presented. Section 3 summarizes the TF-SISR algorithm and presents the TD-based SISR. Section 4 compares the results of the proposed technique against TF-SISR for simulated and real CBCT data, and shows its robustness to noise. Sections 5 and 6 provide a discussion about the results, and give conclusive remarks and perspectives of this work.

\section{Tensor operations}
\label{sec:tensdecomp}

A 3D image can be defined as a third-order tensor $\boldsymbol{X} \in \mathbb{R}^{I\times J\times K}$. Its mode-n fibers are the analogues of columns and rows ($X(:,j,k), X(i,:,k) $ and $ X(i,j,:)$).

$\boldsymbol{X}$ can be multiplied by a 2D matrix along each of its n dimensions ($P_1\in\mathbb{R}^{I^{*}\times I},P_2\in\mathbb{R}^{J^{*}\times J},P_3\in\mathbb{R}^{K^{*}\times K} \quad |$ $ \quad I^{*},J^{*},K^{*}\in\mathbb{Z}$ respectively), called the mode-n products ($\times_n$) \cite{hatvani2018tensor} 
\begin{equation}
\begin{split}
 \boldsymbol{T} = \boldsymbol{X}\times_1 P_1\times_2 P_2\times_3 P_3 =\\
\sum_{i=1}^{I}\sum_{j=1}^{J}\sum_{k=1}^{K} X(i,j,k) P_1(:,i)\circ P_2(:,j) \circ P_3(:,k),\\
\end{split}
\end{equation}

\noindent where $\boldsymbol{T}\in\mathbb{R}^{I^{*}\times J^{*}\times K^{*}}$, and $\circ$ is the outer product.

The tensor unfolding flattens a tensor into a 2D matrix. It can happen along any of the modes; the mode-n fibers are ordered alphabetically as columns of the matrix.

%In case of the CPD this multiplication can be embedded as
%\begin{equation}
%\label{modeprod}
%\boldsymbol{X}\times_1 P_1\times_2 P_2\times_3 P_3 = [\![ P_1 {U^1},P_2{U^2},P_3{U^3} ]\!].
%\end{equation}

\subsection{Canonical Polyadic Decomposition}

The CPD \cite{kolda2009tensor} used in our previous work \cite{hatvani2018tensor} factorizes the tensor $\boldsymbol{X}$ as a sum of $R$ rank-1 tensors (such tensors can be written as the outer product of three 1D arrays, here  $U^n(:,r) | n=1,2,3$)
\begin{equation}
\label{build}
\begin{split}
&\boldsymbol{X} \approx  \sum_{r=1}^R U^1{(:,r)} \circ U^2{(:,r)}, \circ U^3{(:,r)},\\
\end{split}
\end{equation}

\noindent where $\overline{U} =  \left\lbrace {U^1}\in\mathbb{R}^{I\times R}, {U^2}\in\mathbb{R}^{J\times R}, {U^3}\in\mathbb{R}^{K\times R} \right\rbrace$
is a set of three 2D matrices, known as the CPD of the tensor $\boldsymbol{X}$.
The minimal number of such rank-1 tensors that can exactly compose the original tensor is called the rank of the tensor, $F$. The tensor rank is NP-hard to find, but using it a unique decomposition can be achieved under mild conditions \cite{chiantini2012generic}.
The shorthand of this decomposition is  $\boldsymbol{X} = [\![{U^1}, {U^2}, {U^3}]\!]$.

% A tensor might be multiplied by a 2D matrix along each of its n dimensions ($P_1,P_2,P_3$), called the mode-n products ($\times_n$). The mode-n fibers of the tensor are pre-multiplied by the 2D matrix, one at a time. In case of the CPD this multiplication can be embedded as
%\begin{equation}
%\label{modeprod}
%\boldsymbol{X}\times_1 P_1\times_2 P_2\times_3 P_3 = [\![ P_1 {U^1},P_2{U^2},P_3{U^3} ]\!].
%\end{equation}
%Matricization is the process when a 3D tensor is unfolded into a 2D matrix along a selected direction. Using the CPD it can be realized using the Kathri-Rao product ($\odot$)
%\begin{equation}
%\label{matri2}
%\begin{split}
%\boldsymbol{X}^{(1)} = U^1(U^3\odot U^2)^T\\
%\end{split}
%\end{equation}

\subsection{Tucker Decomposition}
\label{TDsec}
The n-rank of a tensor is the number of its independent mode-n fibers. For a third-order tensor three such values, $\{$1-rank, 2-rank, 3-rank$\}$ can be defined. While the CPD defines the rank of a tensor, $F$, for the TD the n-rank is a more practical tool - but the two terms should not be confused. 
TD (also called higher order SVD, or multi-linear SVD) is the multidimensional generalization of the 2D SVD, written as
\begin{equation}
\label{eq:TD}
\boldsymbol{X} \approx \boldsymbol{\Sigma} \times_1 V_1 \times_2 V_2 \times_3 V_3,
\end{equation}

\noindent where $\boldsymbol{\Sigma}$ is a core tensor, and $V_1\in\mathbb{R}^{I\times R_1}, V_2\in\mathbb{R}^{J\times R_2}, V_3\in\mathbb{R}^{K\times R_3}$ are the orthonormal bases of the subspaces spanned by the mode-n fibers \cite{kolda2009tensor}. If $R_1,R_2,R_3$ equal the n-ranks of $\boldsymbol{X}$, the decomposition is unique, but for lower numbers and in the presence of noise it becomes inexact.
 As opposed to the 2D SVD, $\boldsymbol{\Sigma}$ is not diagonal, but shows the level of interaction between the different modes. Due to orthonormality, it can be obtained as
\begin{equation}
\label{eq:G}
\boldsymbol{\Sigma} = \boldsymbol{X} \times_1 V_1^T \times_2 V_2^T \times_3 V_3^T,
\end{equation}
\noindent The shorthand of TD is $\boldsymbol{X} = [\![\Sigma; {V_1}, {V_2}, {V_3}]\!]$.
Singular values similar to the 2D case ($SV_n$ for $n=1,2,3$) can be defined. These are calculated as the Frobenius norms of 2D slices taken from the core tensor $\boldsymbol{\Sigma}$, along mode-n, fixing the index of the component in question.

\section{Methods}
\label{sec:methods}

\subsection{Problem formulation}

The SISR image degradation model considered herein assumes that the HR image $\boldsymbol{X}$ is blurred by a separable (usually Gaussian) kernel $h$, is downsampled by rate $r$, and corrupted by additive white Gaussian noise $\boldsymbol{N}$. The output LR image $\boldsymbol{Y}\in\mathbb{R}^{I/r\times J/r \times K/r}$ is given by the vector-matrix equation
\begin{equation}
    \textnormal{vec}(\boldsymbol{Y}) =  DH \textnormal{vec}(\boldsymbol{X}) + \textnormal{vec}(\boldsymbol{N}),
\end{equation}
\noindent where $H$ is a block-circulant matrix with circulant blocks defined from $h$,  $D$ is an  $r$-fold downsampling operator, and vec$(\cdot)$ vectorizes a tensor alphabetically. In this work the blurring kernel is assumed to be known, and is estimated for the current application as explained in \cite{hatvani2018deep}.

\subsection{Previous TF-SISR using CPD}

In our previous work \cite{hatvani2018tensor} the denoising, deconvolution and upsampling of $\boldsymbol{Y}$ were implemented jointly, separated for the 3 dimensions.
As both the blurring and downsampling operations are separable to the 3 dimensions ($D_1,D_2,D_3$ and $H_1, H_2,H_3$), the alternating least squares algorithm finding the optimal CPD can be realized with
\begin{equation}
\label{solution}
\begin{split}
U^1 = (D_1H_1)^{\dagger}\boldsymbol{Y}^{(1)}(D_3H_3U^3\odot D_2H_2U^2)^{\dagger T}\\
U^2 = (D_2H_2)^{\dagger}\boldsymbol{Y}^{(2)}(D_3H_3U^3\odot D_1H_1U^1)^{\dagger T}\\
U^3 = (D_3H_3)^{\dagger}\boldsymbol{Y}^{(3)}(D_2H_2U^2\odot D_1H_1U^1)^{\dagger T},\\
\end{split}
\end{equation}

\noindent as derived in \cite{hatvani2018tensor}. These steps are repeated until $\overline{U}$ converges. $\boldsymbol{Y^{(n)}}$ is the n-unfolded tensor, and $\odot$ also implements unfolding, only in the CPD-domain. Denoising is realized by choosing a small enough $R$. The symbol $\dagger$ denotes the Moore-Penrose pseudoinverse with Tikhonov regularization
\begin{equation}
\label{eq:tikh}
    A^\dagger = (A^TA + \epsilon I)^{-1}A^T.
\end{equation}

\subsection{Proposed TD-SISR}

The idea of the method proposed herein is to denoise the image before deconvolution, in order to stabilize the ill-posed operation, as earlier suggested in the literature \cite{denoise_deconv}.

As explained in Section \ref{TDsec}, the singular values of each mode ($SV_1,SV_2,SV_3$) can be calculated from $\boldsymbol{\Sigma}$. Similarly to the 2D case, by picking the relevant components having a singular value higher than a threshold $R_n$, a denoised version of $\boldsymbol{Y}$, $\boldsymbol{\hat{Y}}$ may be achieved \cite{denoise_svd}.
\begin{equation}
\label{eq:G}
\begin{split}
& \boldsymbol{\hat{Y}} = \overline{\overline{\Sigma}} \times_1 \overline{V_1} \times_2 \overline{V_2} \times_3 \overline{V_3}\\,
\textnormal{where}& \\
    & \overline{V_n} = V_n(:,i) \quad | \quad SV_n(i) \geq R_n \\
    & \boldsymbol{\overline{\Sigma}} = \boldsymbol{Y} \times_1 \overline{V_1}^T \times_2 \overline{V_2}^T \times_3 \overline{V_3}^T \\
\end{split}
\end{equation}

Unlike in the 2D case, this truncated approximation might not be optimal in the least squares sense, but gives a reasonable estimate \cite{kolda2009tensor}.

After obtaining the denoised image, the deconvolution is realized using a Tikhonov-regularized deconvolution separated for the three modes.
\begin{equation}
    \boldsymbol{\hat{X}} = \boldsymbol{\hat{Y}} \times_1 (D_1H_1)^\dagger \times_2 (D_2H_2)^\dagger \times_3 (D_3H_3)^\dagger .
\end{equation}

\section{Results}
\label{sec:result}

\subsection{Datasets and metrics}

HR dental images were acquired using a QuantumFX  micro-CT  system  (Perkin  Elmer, resolution  1  LP/mm  at  50\% MTF, meaning that spatial frequencies of 1 line pair per mm are depicted at 50\% of the modulation transfer function). In simulation these images were blurred (with a Gaussian kernel, standard deviation $\sigma_1$=$\sigma_2$=$\sigma_3$=8), downsampled (at $r$=2) and Gaussian noise was added (white noise at different SNR levels). For simulations, a lower premolar was chosen (280×268×492 pixels). For real data experiments the LR images were CBCT images acquired with a Carestream 81003D system (resolution 10  LP/mm  at  50\% MTF). Following \cite{hatvani2018deep}, a Gaussian blurring kernel was estimated from the CBCT data, and served as an input for the SISR algorithms. For this application, the estimated standard deviations of the Matlab Gaussian functions were $\sigma_1$=8.2, $\sigma_2$=7.5, $\sigma_3$=1.3. The volumes naturally contain noise because of the measurements and of the reconstruction algorithm. However, the level of noise is low for extracted teeth due to the absence of surrounding structures. To mitigate this and to further explore the robustness of the SISR algorithms to noise, Gaussian noise corresponding to different SNRs was artificially added to the experimental data. The source of the real data was an upper molar (324×248×442 pixels).

The improvement of the LR images was measured through their peak signal-to-noise ration (PSNR) and structural similarity index (SSI) \cite{wang2004image}, calculated against the micro-CT HR images (which are considered as the ground truth due to their high spatial resolution and SNR, but unavailable in clinics because of the excessive radiation dose).
%The added white noise introduced negative values in the background, causing different shifts of the dynamic range at different noise levels. In order to acquire comparable metrics, they
The metrics were calculated for the teeth only, excluding the background. 

The efficiency of the algorithms was also measured on the segmented root canals following a dedicated method developed in \cite{michetti2017comparison}, calculating their volume and Dice coefficient \cite{dice1945measures} against the segmented HR micro-CT volumes.  

The algorithms were implemented in Matlab 2017 \cite{MATLAB:2017}, and for basic tensor operations Tensorlab \cite{tensorlab30} was used. 

\begin{figure*}[t!]
    \centering
    \includegraphics[width=18cm]{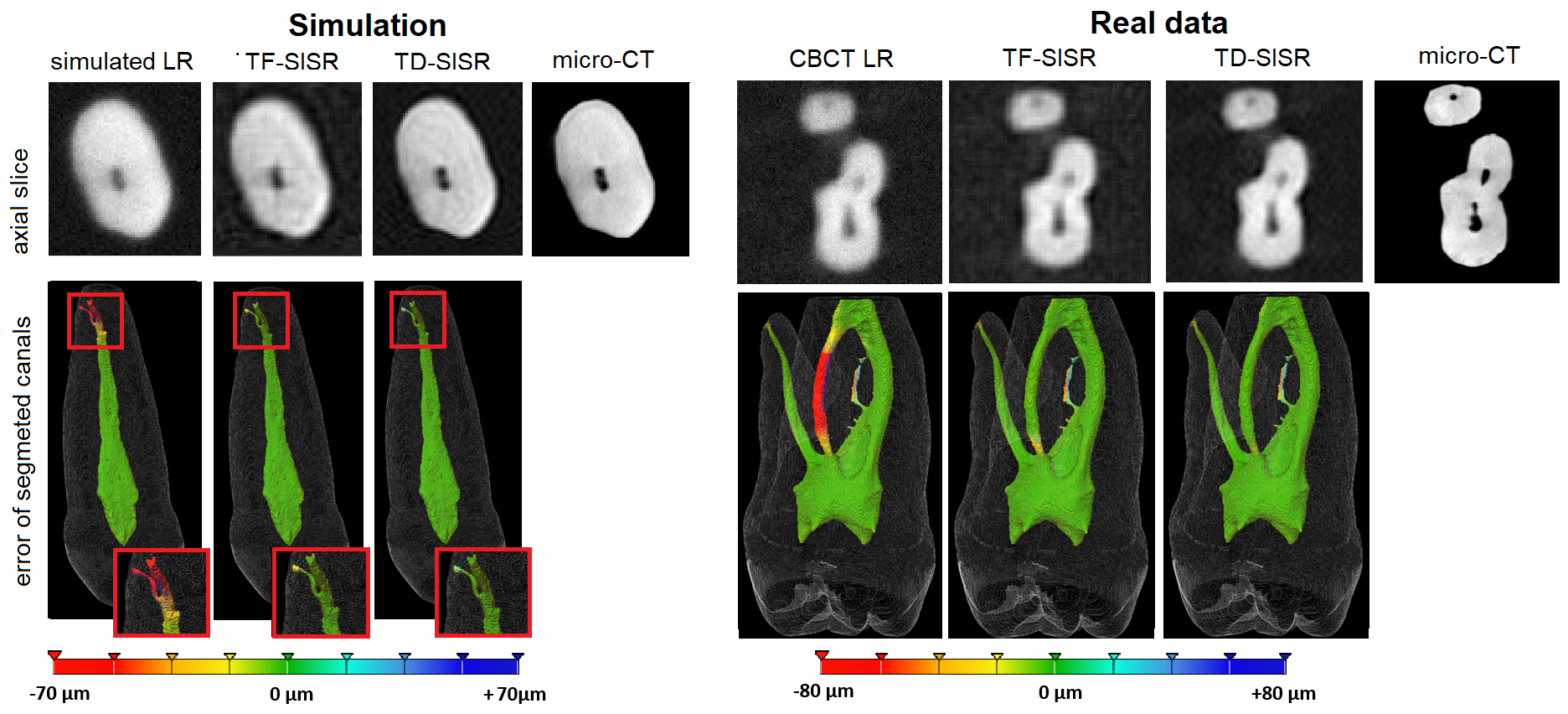}
    \caption{Results of SISR methods under 25 dB noise, both in simulation and in real data. The first row shows a single axial slice taken from the volumes. The second row shows the distance between the segmented HR and LR, enhanced LR volumes.}
    \label{fig:qualitative}
\end{figure*}

\subsection{Simulation results}

\begin{figure}[t!]
    \centering
    \includegraphics[width=8.8cm]{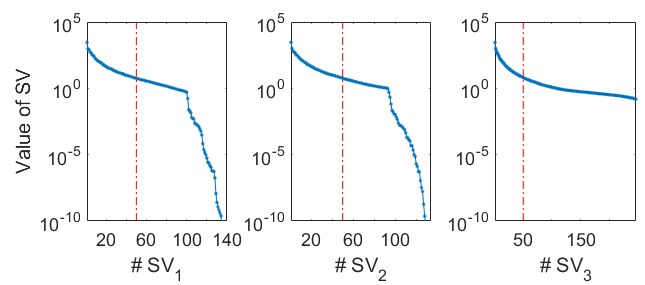}
    \caption{Simulation - singular values without added noise in all three modes, on a logarithmic scale. The vertical lines represent the chosen truncation thresholds}
    \label{fig:SV_sim}
\end{figure}

For TF-SISR $R=500$ was chosen, and in both methods $\epsilon=1$ was set for (\ref{eq:tikh}) following \cite{hatvani2018tensor}. As it can be seen in Fig. \ref{fig:SV_sim}, the singular values decay rapidly (mind the logarithmic scale). For TD-SISR $R_1$=$R_2$=$R_3$=40 (SV values under 1) were chosen as they were generally sufficient for all noise levels.

Table \ref{tab:sim} shows the quantitative results of the SISR methods. The PSNR is improved for each case and both methods, compared to the simulated LR image. TD-SISR gave better results, except for the extremely noisy, 20 dB case.
The SSI gave similar results, with no improvement in the 20 dB case.
After these results the segmentation was carried out at 25 dB.
%The volumes of the canals ($mm^3$, LR or enhanced LR over HR) show that both method estimated the volumes better than from the LR image.
The improvement is confirmed by the Dice coefficients, showing the superiority of the TD-SISR method.

\begin{table}[h!]
\centering
\renewcommand{\arraystretch}{1.3}
\caption{Metrics in simulation}
\label{tab:sim}
\small
\begin{tabular}{l|ccc}
\hline
&Simulated LR&  TF-SISR& TD-SISR\\
\hline\hline
runtime&-& 17.96 s & \textbf{1.86 s} \\
\hline
&&PSNR (dB)&\\
\hline
no noise & 28.56 & 31.48 & \textbf{34.99} \\
30 dB & 28.45 & 31.17 & \textbf{34.39}\\
25 dB & 28.36 & 31.08 & \textbf{31.40} \\
20 dB & 27.98 & \textbf{30.01} & 29.33 \\
\hline
&&SSI [0, 1]&\\
\hline
no noise & 0.9623 & 0.9680& \textbf{0.9823}  \\
30 dB & 0.9612 & 0.9650 & \textbf{0.9763} \\
25 dB & 0.9572 & 0.9595 & \textbf{0.9653} \\
20 dB & 0.\textbf{9463} & 0.9453 & 0.9417 \\
\hline
&\multicolumn{3}{c}{Segmentation at 25 dB}\\
\hline
%volume ratio & 7.19$/$8.68 & 8.44$/$8.68 & 8.51$/$8.68 \\
Dice & 0.8976 & 0.9242& \textbf{0.9425}  \\
\hline
\end{tabular}
\end{table}

\subsection{Real data results}

\begin{figure}[t!]
    \centering
    \includegraphics[width=8.2cm]{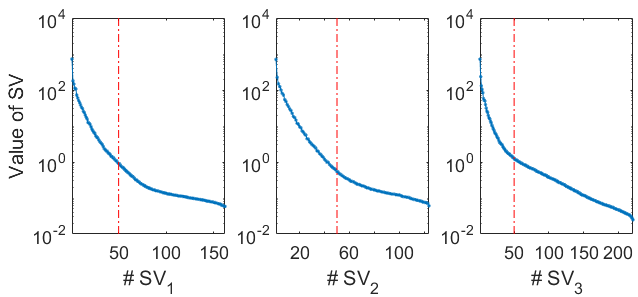}
    \caption{Real data - singular values without added noise in all three modes, on logarithmic scale. The vertical lines represent the chosen truncation thresholds}
    \label{fig:SV_real}
\end{figure}

For the real data in TF-SISR the same settings were used as in simulation, and for TD-SISR $R_1,R_2,R_3 = 50$ was set after plotting the SVs (Fig. \ref{fig:SV_real}).

The metrics have shown milder improvements compared to the simulation, but both PSNR and SSI improved in all cases with both methods, and TD-SISR gave superior results. The volume of the segmented 25 dB images also improved regarding the Dice coefficient.

\begin{table}[h!]
\centering
\renewcommand{\arraystretch}{1.3}
\caption{Metrics in simulation}
\label{tab:real}
\small
\begin{tabular}{l|ccc}
\hline
&Simulated LR&  TF-SISR& TD-SISR\\
\hline\hline
runtime&-& 17.71 s &  \textbf{1.46 s}\\
\hline
&&PSNR (dB)&\\
\hline
no noise & 19.55 & 21.25 & \textbf{21.61} \\
30 dB & 19.30 & 20.84 & \textbf{21.57}\\
25 dB & 19.10 & 20.13 & \textbf{21.09} \\
20 dB & 18.91 & 20.21 & \textbf{20.29} \\
\hline
&&SSI [0, 1]&\\
\hline
no noise & 0.8647  & 0.8907 & \textbf{0.8935}  \\
30 dB & 0.8610 & 0.8870 & \textbf{0.8929} \\
25 dB & 0.8478 & 0.8784 & \textbf{0.8908} \\
20 dB & 0.8173 & 0.8555 & \textbf{0.8814} \\
\hline
&\multicolumn{3}{c}{Segmentation at 25 dB}\\
\hline
%volume ratio & 30.88$/$37.28& 33.00$/$37.28 & 33.16$/$37.28  \\
Dice & 0.8939&0.9189& \textbf{0.9304}  \\
\hline
\end{tabular}
\end{table}

\section{Discussion}
In contrast to the earlier TF-SISR method no iterations are applicable in TD-SISR. In TD-SISR 3 thresholds $R_1,R_2,R_3$ have to be defined for the three modes, while in TF-SISR only one parameter, $R$ influences the denoising step. However, the singular values of TD-SISR correspond to the importance of the components, while $R$ in TF-SISR bears no such meaning. This makes the setting of TD-SISR parameters easier, and its efficiency is validated by the qualitative and quantitative results. The runtime of TD-SISR is lower because of the lack of iterations, but calculating the SVD for even larger volumes might be a bottleneck \cite{svd}.

\section{Conclusion}
\label{sec:conc}

In this paper a new SISR technique was proposed, using Tucker decomposition for the denoising, and a Tikhonov-regularized deconvolution. Even though 2 additional parameters have to be set, it gave faster and quantitatively better results in noisy images compared to our previous method, TF-SISR. Images of 280×268×492 and 324×248×442 pixels were super-resolved under 2 s with standard Matlab implementation. In future work, the connection between the TF-SISR and TD-SISR parameters along with their robustness, and the general inverse problem including thresholding constraint will be investigated.

% Below is an example of how to insert images. Delete the ``\vspace'' line,
% uncomment the preceding line ``\centerline...'' and replace ``imageX.ps''
% with a suitable PostScript file name.
% -------------------------------------------------------------------------
%\begin{figure}[htb]

%\begin{minipage}[b]{1.0\linewidth}
%  \centering
%  \centerline{\includegraphics[width=8.5cm]{img/image1}}
%  \vspace{2.0cm}
%  \centerline{(a) Result 1}\medskip
%\end{minipage}
%
%\begin{minipage}[b]{.48\linewidth}
%  \centering
%  \centerline{\includegraphics[width=4.0cm]{img/image3}}
%  \vspace{1.5cm}
%  \centerline{(b) Results 3}\medskip
%\end{minipage}
%\hfill
%\begin{minipage}[b]{0.48\linewidth}
%  \centering
%  \centerline{\includegraphics[width=4.0cm]{img/image4}}
%  \vspace{1.5cm}
%  \centerline{(c) Result 4}\medskip
%\end{minipage}
%
%\caption{Example of placing a figure with experimental results.}
%\label{fig:res}
%
%\end{figure}
%

%\vfill
%\pagebreak

\bibliographystyle{IEEEbib}
\bibliography{strings,refs}

\end{document}